\documentclass[final,5p,times,twocolumn]{elsarticle}

\usepackage{graphicx}%
\usepackage{multirow}%
\usepackage{amsmath,amssymb,amsfonts}%
\usepackage{amsthm}%
\usepackage{mathrsfs}%
\usepackage[title]{appendix}%
\usepackage{xcolor}%
\usepackage{textcomp}%
\usepackage{manyfoot}%
\usepackage{booktabs}%
\usepackage{algorithm}%
\usepackage{algorithmicx}%
\usepackage{algpseudocode}%
\usepackage{listings}%

\begin{document}
\begin{frontmatter}

\title{Effect of Cr Segregation on Grain Growth in Nanocrystalline $\alpha$-Fe Alloy: A Multiscale Modelling Approach}

\author[a,b]{Sandip Guin}
\author[a]{Albert Linda}
\author[b,c]{Yu-Chieh Lo}
\author[a]{Somanth Bhowmick*\corref{cor}}
\ead{bsomnath@iitk.ac.in}
\author[a]{Rajdip Mukherjee *\corref{cor}}
\ead{rajdipm@iitk.ac.in}

\address[a]{Department of Materials Science and Engineering, Indian Institute of
Technology, Kanpur, Kanpur-208016, UP, India}
\address[b]{International College of Semiconductor Technology. International College of Semiconductor Technology, Hsinchu, 300, Taiwan}
\address[c]{Department of Materials Science and Engineering, International College of Semiconductor Technology, Hsinchu, 300, Taiwan}

\begin{abstract}
We present a multiscale modelling framework that integrates density functional theory (DFT) with a phase-field model (PFM) to explore the intricate dynamics of grain growth in nanocrystalline $\alpha$-Fe single-phase alloy in the presence of chromium (Cr) segregation. We begin our study by validating our simulation results for equilibrium segregation in stationary GB with Mclean isotherm. 
Polycrystal simulations featuring nanocrystalline grains at different temperatures reveal that the grain growth kinetics depends on the ratio of Cr diffusivity to intrinsic GB mobility. 
In the absence of segregation, the relationship between the square of average grain size ($d^2$) and time (t) demonstrates a linear correlation. We observe that the $d^2$ vs. $t$ plot exhibits a consistent linear trend up to a threshold grain size, independent of Cr segregation at GB.
However, when Cr is segregated at GB, a deviation from this linear trend with a decreasing slope is evident within the temperature range of 700K to 900K beyond the threshold size. This threshold grain size decreases with increasing temperature. Notably, at 1000K, the deviation from the linear trend is observed from the initial stages of grain growth with segregation, albeit with a linear trend exhibiting a smaller slope. We also present an analytical formulation based on Cahn solute drag theory to predict grain growth behaviour in the presence of solute segregation and our simulation results well aligned this analytical formulation.  
\end{abstract}


\end{frontmatter}

\section{Introduction}\label{sec1}
Nanocrystalline materials contribute significantly to enhanced material strength~\cite{WU2023106083, GLEITER1989223, BU2023171238, HAMDI2023170826}. One can understand the correlation between grain size and yield strength using the Hall-Petch relationship~\cite{ DONG2023168515, PENG2023168008, HAMDI2023170826}. The increase of strength for nanocrystalline materials arises from the increased density of grain boundaries (GB), where the interactions between dislocations are impeded, hindering their movement and resulting in improved mechanical properties~\cite{JINLONG2023167791, PENG2023168008, BU2023171238}. For this reason, preparing nanocrystalline materials remains one of the most commonly used strengthening mechanisms. For example, Qin \emph{et al.} showed that Ni-Co alloy synthesized by electrodeposition, with an average grain size of 15 nm, demonstrates high strength (1920–2250 MPa) and improved ductility (10–13\%) in room temperature uniaxial tensile test~\cite{QIN2010S439}.

However, nanocrystalline materials have a substantial volume fraction of GB~\cite{DRIVER2004819}. Minimizing the surplus grain boundary energy is the primary driving force behind grain growth in these nanocrystalline materials, resulting in the loss of their nanocrystalline characteristics~\cite{GERTSMAN1994577, AMES20084255}. This leads to poor thermal stability of nanocrystalline materials~\cite{GLEITER20001}. Stabilizing grain growth in nanocrystalline materials using GB solute segregation is a promising approach in materials science to enhance the stability and control of the microstructure~\cite{D0NR07180C, MILLETT20072329, WEISSMULLER1993261}. GB solute segregation is a phenomenon where solutes accumulate at the boundaries between individual crystalline grains within a solid material~\cite{LEJOEK2016,YE2021102808}. GB segregation stabilizes nanocrystalline grain structure by reducing GB energy and slowing down the grain growth due to solute drag effect~\cite{Lejcek201783, Raabe2014253, MILLETT20072329, KIRCHHEIM2002413, HILLERT1976731, CAHN1962789, CANTWELL20141}. For example, Wang \emph{et al.} showed that nickel-based alloy with nanocrystalline structure and co-segregation of multiple solutes exhibited good thermal stability when subjected to high-temperature annealing~\cite{WANG2021158326}. Similarly, Darling \emph{et al.} studied the stabilization of nanocrystalline FeZr alloy where both solute drag and reduction in GB energy are stabilization mechanisms~\cite{DARLING2008530}.

Quantifying GB segregation and its effect on drag force is not straightforward. There is a complex coupling between GB solute segregation and different material parameters such as the segregation energy, diffusivity of solute, intrinsic GB mobility, temperature, as well as the GB velocity due to grain growth~\cite{D0NR07180C, HEO20117800, CAHN1962789, WANG2023167717, MONDAL2014206, ITO2024107849, MA2020101388}. It is well known that GB segregation increases with increasing segregation energy~\cite{mclean1958grain}. The diffusion of solute significantly influences GB segregation during the migration of GB~\cite{KAUR2020109685, YE2021102808}. When the solute has a higher diffusivity, it can more easily catch the migrating GB, leading to increased segregation compared to situations with lower diffusivity. Cahn \emph{et al.} explored the correlation between the extent of GB segregation and GB velocity, highlighting the dependency of the solute drag force on this velocity~\cite{CAHN1962789}. As the GB velocity increases, segregation decreases, resulting in a lower drag force. On the other hand, a stationary GB exhibits equilibrium segregation at a constant temperature relative to the bulk solute concentration~\cite{CAHN1962789, mclean1958grain}. With an elevation in temperature, there is a corresponding reduction in the extent of equilibrium segregation at the GB~\cite{mclean1958grain}. 

In nanocrystalline materials, the substantial driving force for grain growth results in high GB velocity, yielding lower GB segregation and negligible drag force. During grain growth, the driving force diminishes with increasing grain size, causing a reduction in GB velocity, heightened GB segregation, and increased drag force~\cite{GRONHAGEN2007955}. For example, Zhao \emph{et al.} showed that with increasing grain size, phosphorus segregation at GB increases in steel~\cite{Zhao_2017}.
Therefore, in the early stages of grain growth in nanocrystalline materials, the level of GB segregation tends to be minimal due to the high velocity of the GB, resulting in negligible drag force. As time progresses and grain growth proceeds, the average grain size increases, leading to increased segregation at the GB. Eventually, a threshold grain size is reached, from which a significant amount of segregation occurs at the GB, resulting in a notable drag force on the moving GB. Subsequently, grain growth rate reduction assisted by drag force begins to take effect beyond that threshold grain size.
Consequently, considering the presence of GB segregation, a comprehensive investigation into the kinetics of grain growth in nanocrystalline materials becomes imperative. Specifically, understanding the threshold grain size at which solute drag force becomes significant is crucial for anticipating and managing the stabilization of nanocrystalline grain structure using GB segregation.

This work presents a multiscale modeling framework, combining  Density Functional Theory (DFT) and phase-field simulations, to investigate the grain growth phenomena in nanocrystalline single-phase $\alpha$-Fe alloy in the presence of chromium (Cr) segregation. DFT had already been used to study various properties of GB\cite{MAI2022117902, HU2020109271, 10.1063/1.4867400}, including GB energies \cite{ZHENG202040, 10.1063/1.4867400}, work of separation \cite{WANG2016279}, and segregation energies \cite{RAZUMOVSKIY2018122}. These simulations have exhibited good qualitative agreement compared to experimental data \cite{EBNER2021117354}. Grain growth study is carried out using phase field simulation, which is a computational technique used in materials science to model and study the evolution of microstructure and phase transition in complex systems~\cite{CHA20023817, LiWangYang+2010+555+559, GRONHAGEN2007955}. It provides a powerful and versatile approach for simulating various phenomena, including solidification, crystal growth, phase separation, GB segregation, and other phase transitions~\cite{KIM2016150, VERMA2020155163, GUO2023105811, LVOV2023106209}. Phase-field simulations are particularly valuable because they can capture intricate details of microstructural evolution over time and space, making them indispensable tools for understanding and designing materials and processes. Researchers have extensively used phase-field simulation to study grain growth and GB segregation in the last few decades. For example, Wu \emph{et al.} studied abnormal grain growth in Mg alloy using phase-field modelling~\cite{WU2020100790}. 
Heo \emph{et al.} studied the effect of elastic misfit strain on GB segregation~\cite{HEO20117800}. Kundin et al. studied the grain growth with segregation in ceramic matrix mini-composites using phase-field modelling~\cite{KUNDIN2021110295}.

In this paper, we study the effect of Cr segregation on the grain growth behaviour of single phase $\alpha$-Fe. We consider a  $\sum3(1 \bar{1} 1)[110]$ GB, generated within a periodic cell \cite{MAI2022117902} and calculate the segregation energy of Cr using DFT simulation. The segregation energy, GB energy and GB width obtained from DFT calculation is used as an input parameter for phase-field simulation. Initially, we studied the equilibrium GB segregation in stationary GB using bi-crystal simulation, followed by nanostructured polycrystalline simulation, to understand the impact of Cr segregation on grain growth at different temperatures.


\section{Model}\label{sec2}

\begin{figure}[htpb]
\centering 
\includegraphics[width=1\linewidth]{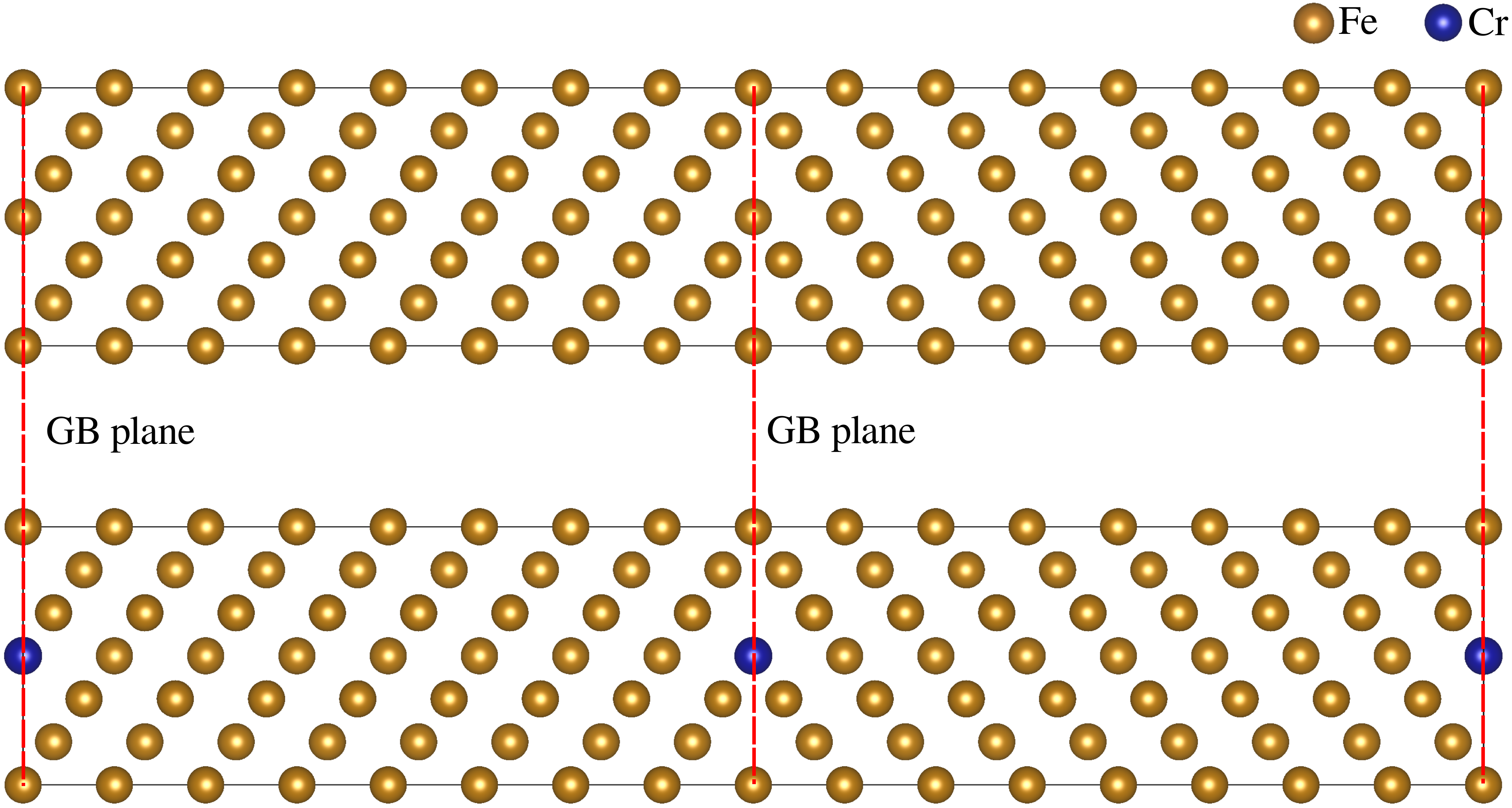}
\caption{Atomistic structure (generated using VESTA\cite{VESTA}) of a GB without (top) and with (bottom) Cr segregation.}
\label{fig:atomic_seg}
\end{figure}

\subsection{Estimation of segregation energy}
We generate a high-angle tilt GB ($\sum3(1 \bar{1} 1)[110]$) using \textit{aimsgb} tool\cite{CHENG201892}. The resultant supercell has dimensions of $39.63\AA$, $7.00\AA$, and $4.04\AA$ along the three repeat directions, and it contains two grain boundaries, separated by a distance of $19.81\AA$ [Figure~\ref{fig:atomic_seg}]. GB energy is just the energy difference per area between between bulk and GB structure\cite{lejcek2010grain} which is given by:
\begin{equation}
    \sigma_{gb} = \frac{E^{GB}-E^{bulk}}{2A}.
    \label{eq:gb_energy}
\end{equation}
Here, $E^{GB}$ and $E^{bulk}$ represent the energies of the supercells with and without the GB, respectively, while $A$ is the grain boundary area. The factor of two in the denominator accounts for two GBs per supercell. The segregation energies are calculated using the following equation\cite{lejcek2010grain}:
\begin{equation}
    E_{seg.} = \frac{(E_{solute}^{GB}-E_{solute}^{bulk})-(E^{GB}-E^{bulk})}{2}.
\end{equation}
Here, $E_{solute}^{GB}$ and $E_{solute}^{bulk}$ denote the energies of the supercell with solute substitution in the presence and absence of the GB, respectively. 

\subsection{Phase Field Simulation}
\begin{figure}[ht]
\centering 
\includegraphics[width=0.70\linewidth]{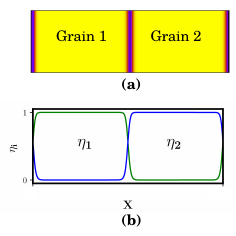}
\caption{(a) Illustration of a bi-crystal system featuring a flat grain boundary with finite thickness. (b) Graph depicting the variation of the grain order parameter $\eta_i$, where $\eta_1$ corresponds to grain 1 and $\eta_2$ corresponds to grain 2.}
\label{fig:schematic}
\end{figure}
This study expands upon the phase-field model introduced by Heo \emph{et al.} adjusting the parameter governing the GB and solute interaction potential~\cite{HEO20117800}.
Our phase-field methodology utilizes a conserved order parameter denoted as $c$ to depict solute concentration and a set of non-conserved parameters $\eta_i; (i = 1 \dots N)$ for each grain~\cite{HEO20117800, CAHN1962789}. Total free energy ($F_{total}$) is given as:
    \begin{equation}
          \begin{aligned}
            F_{total} = \int_{V}\Big[f_{chem} +c\cdot E  + {\omega} \cdot g({\eta_1},{\eta_2},..,{\eta_N}) + \frac{\kappa_c}{2}|\nabla c|^2 + \\  \sum_{i}\frac{\kappa_{\eta}}{2}|\nabla {\eta_i}|^2 \Big]dV.
          \label{eq:seg_2}
          \end{aligned}
    \end{equation}
The chemical free energy density is denoted by $f_{chem}$. $E$ signifies the interaction of solute atoms (Cr) with the GB. The function ${g({\eta_1},{\eta_2},..,{\eta_N})}$ describes the free energy density of grain structure, while ${\omega}$ represents the height of ${g({\eta_1},{\eta_2},..,{\eta_N})}$. ${\kappa_c}$ and ${\kappa_{\eta}}$ represent the gradient energy coefficients related to the composition ${c}$ and the grain order parameters ${\eta_i}$, respectively. We employ a model based on an ideal solution framework to establish ${f_{chem}}$ in a binary system, articulated as:
\begin{equation}
    \begin{aligned}
    f_{chem} = k_BT[clnc+(1-c)ln(1-c)]. 
    \end{aligned}
    \label{eq:finc}
    \end{equation}
Here,  ${k_B}$ represents the Boltzmann constant, while ${T}$ denotes the temperature. The formula describing ${g({\eta_1},{\eta_2},..,{\eta_N})}$ is as follows:
\begin{equation}
          \begin{aligned}
          g({\eta_1},{\eta_2},....,{\eta_N}) = &0.25+\sum_{i} \Big (-\frac{{\eta_i}^2}{2}+\frac{{\eta_i}^4}{4} \Big )+ \\ & \gamma \sum_{i}\sum_{j>i}{{\eta_i}^2}{{\eta_{j}}^2}.
          \end{aligned}
          \label{eq:fan_chen}
\end{equation}
In this context, the phenomenological constant ${\gamma}$ represents the interactions within the grain order parameter and ${E}$ is defined as $-m.\omega.g({\eta_1},{\eta_2},..,{\eta_N})$. The parameter ${m}$ dictates the degree of influence of the interaction between solute and GB.

Throughout the evolution process, we numerically solve the Cahn-Hilliard equation~\cite{CAHN1961795} for the conserved parameter ($c$) and the Allen-Cahn equation~\cite{ALLEN19791085} for the non-conserved parameter (${\eta_i}$), as described by Equation~\ref{eq:seg_5} and Equation~\ref{eq:seg_6}, respectively.
\begin{equation}          
          \frac{\partial c}{\partial t} = \nabla\cdot M_c \nabla \Big (\frac{\partial f_{chem}}{\partial c} -m\cdot\omega\cdot g-\kappa_c{\nabla}^2c \Big ),   
          \label{eq:seg_5}
\end{equation}
\begin{equation}          
          \frac{\partial \eta_i}{\partial t} = -L \Big (\omega\frac{\partial g}{\partial \eta_i}-m{\cdot}c\cdot\omega\cdot\frac{\partial g}{\partial \eta_i}  -\kappa_{\eta}{\nabla}^2\eta_i \Big ).   
          \label{eq:seg_6}
\end{equation}
Here, ${M_c}$ stands for the solute's mobility, ${L}$ corresponds to the GB relaxation parameters, and ${t}$ represents time. The mobility factor ${M_c}$ assumes the form of ${[M^o_c \cdot c(1-c)]}$, where ${M^o_c=\frac{D}{k_BT}}$, with ${D}$ representing the diffusivity of Cr in $\alpha$-Fe. The kinetic coefficient ($L$) bears a relationship to the GB  mobility ($M$) of $\alpha$-Fe, as outlined in the work by Heo \emph{et al}~\cite{HEO20117800}:
\begin{equation}          
           L=\frac{M\sigma_{gb}}{\kappa_{\eta}},   
          \label{eq:multi_L_1}
\end{equation}
where $\sigma_{gb}$ is the GB energy. In this context, it is worth noting that both $D$ and $M$ exhibit temperature-dependent variations that conform to the Arrhenius-type equation, given by: $D = D_o \exp\left(-\frac{Q_D}{RT}\right)$ and $M = M_o \exp\left(-\frac{Q_M}{RT}\right)$~\cite{fazeli2005application,1990255}. Here, $D_o$ and $M_o$ is a constant. $Q_D$ and $Q_M$ are the activation energy for diffusion and GB mobility, respectively. 
In the Cahn-Hilliard equation, we do not solve the fourth-order term by assuming the gradient energy coefficient in composition, $\kappa_c$, to be zero. Hence the Equation~\ref{eq:seg_5} becomes a simple diffusion equation.

In our phase field model, the interaction between solute and GB is defined by an interaction potential ($E$) given by $-m\omega.g({\eta_1},{\eta_2},..,{\eta_N})$. Here, $g({\eta_1},{\eta_2},..,{\eta_N})$ represents the free energy density of grain structure, while $\omega$ is constant. In our specific case, $\omega$ has a fixed value, and $g({\eta_1},{\eta_2},..,{\eta_N})$ is zero in the bulk region, gradually increasing within the GB and reaching a maximum value at the centre of GB. Consequently, the only variable we can adjust is $m$, which allows us to control the strength of the solute interaction at the grain boundary and thus govern the extent of solute segregation. Heo \emph{et al.}~\cite{HEO20117800} have elucidated the relationship between the interaction potential ($E$) and segregation energy as follows:
\begin{equation}          
           -m.\omega .g|_{\text{at Centre of GB}} = E_{seg}. 
          \label{eq:multi_L}
\end{equation}
\textcolor{black}{In this scenario, $g|_{\text{at Centre of GB}}$ represents the value of $g({\eta_1},{\eta_2},....,{\eta_N})$ at the centre of the GB, as defined by the phase-field approach.
In phase-field modelling, the GB possesses a finite thickness, with its centre point situated at the midpoint of this finite region. In this case, the value of $g|_{\text{at Centre of GB}}$ is 0.0834.} 
Using the Equation~\ref{eq:multi_L}, we can determine the value of $m$, which we use as an input parameter for our phase-field simulation in terms of interaction energy ($E$) in Equation~\ref{eq:seg_2}. Note that the interaction energy term ($E$) already contains the effect of strain energy due to mismatch in atomic diameters between the solute atom and the matrix atom (through DFT calculations). Hence, the strain energy term is not separately included in the total free energy term described in equation~\ref{eq:seg_2}.

\section{Simulation details}\label{sec3}
Atomic and electronic ground states are obtained via DFT calculations, as implemented in Vienna Ab initio Simulation Package (VASP)~\cite{PhysRevB.54.11169}, using an energy cutoff of $400$ eV and Methfessel-Paxton~\cite{PhysRevB.40.3616} electron smearing, with a width of $0.1$ eV. We use energy convergence criteria of $10^{-6}$ eV for the electronic step and force convergence criteria of  $0.01$ eV/\AA~ for the ionic step. We use a k-point mesh of $2\times10\times17$ for Brillouin zone sampling. We use spin-polarized calculations to take into account the ferromagnetic ground state of Fe.

\begin{table}[ht]
\centering
\caption{Simulation parameter}
\begin{tabular}[t]{lcc}
\hline
Parameters&Values\\
\hline
$l_{gb}$  & 0.84 nm\\
$\sigma_{gb}$  & 1.58 $J/m^2$\\
$E_{seg}$  & -0.18 ev\\
$D_o$  &  $3.67\times10^{-3}$ $m^2/s$ ~\cite{1990255}\\
$Q_D$    &  267.4  $KJ/mol$~\cite{1990255}\\
$M_o$    &  5.8  $cm$ $mol/J s$~\cite{fazeli2005application}\\
$Q_M$    &  140  $KJ/mol$~\cite{fazeli2005application}\\
$\kappa_{\eta}$ & $2.25\times10^{-10}$ $Jm^{-1}$\\ 

 $\mu^{o}_{h}$ & $1.08
\times10^9$ $Jm^{-3}$\\ 

 $\omega$ & $1.97\times10^{8}$ $Jm^{-3}$\\
 $\gamma$ & $1.0$\\
$m$ & $1.75\times10^{-27}$ $m^3$\\

\hline
\end{tabular}
\label{Tab:fe_cr_1}
\end{table}%
Our phase-field study begins by conducting a Bi-crystal simulation to examine the equilibrium grain boundary segregation at different temperatures. For bi-crystal simulation, the total simulation domain size is $44.8 \times 12.8$ $nm^2$. In the Bi-crystal simulation, we select two grains with flat GBs (at the middle of the domain) to ensure that the boundary remains stationary over time (shown in Figure~\ref{fig:schematic}(a)). 

Subsequently, we shift our focus to a nanocrystalline simulation to investigate the impact of segregation on grain growth across a range of temperatures. In the polycrystalline simulations, we utilize a domain size of $1600\times1600$ $nm^2$, with an initial set of 2000 grains. The initial average grain size is 35.77 nm, which makes it a nanocrystalline microstructure. We select four distinct temperatures (700K, 800K, 900K, and 1000K) to investigate the segregation behavior at GB. For polycrystalline simulation, the initial Cr concentration is 5 at\%.

The proposed model is employed in the MOOSE software, a finite element method (FEM) based solver~\cite{lindsay2022moose,schwen2023phasefield}. Periodic boundary conditions are applied in both the X and Y directions. We utilize the Newton solver in MOOSE without adaptive meshing and time stepping. It is important to note that the total composition within the entire domain remained conserved throughout the simulation. We use the Grain Tracker algorithm to reduce the non-conserve order parameter for simulating polycrystalline grains structure~\cite{lindsay2022moose,schwen2023phasefield}. All the simulation parameters are provided in Table~\ref{Tab:fe_cr_1}.


\section{Results}\label{sec4}
\subsection{Segregation energy and GB width}
We use DFT to estimate both $E_{seg}$ and $l_{gb}$ required for the phase-field simulation. Initially, we calculate the $\sum3(1 \bar{1} 1)[110]$ GB energy for pure $\alpha$-Fe. Our estimate of 1.58 J/m$^2$ is very close to the previously reported value of 1.56 J/m$^2$~\cite{MAI2022117902}, which validates the reliability of the input parameters for the DFT run. Following this, introducing Cr at the GB plane [Figure~\ref{fig:atomic_seg}] resulted in an estimated $E_{seg}$ value of $-0.18$ eV. The negative sign for $E_{seg}$ indicates a preference for the solute atom to segregate at the GB plane. We evaluate segregation energies at various atomic sites proximal to the GB plane. At approximately $0.42$ nm away from the GB plane, the segregation energy value becomes negligible ($-0.007$ eV, which is $1/25^{th}$ of the value obtained at the GB plane). Based on this observation, we take the GB width ($l_{gb}$) as $0.84$ nm and the segregation energy ($E_{seg}$) as $-0.18$ eV for the phase-field simulation.

After calculating the $E_{seg}$ value using DFT calculation, we calculate the value of the parameter $m$ using Equation~\ref{eq:multi_L}. The value of $m$ for our simulation is $1.75\times10^{-27}$ $m^3$.

\subsection{Bi-crystal system}
\begin{figure}[ht]
\centering 
\includegraphics[width=0.75\linewidth]{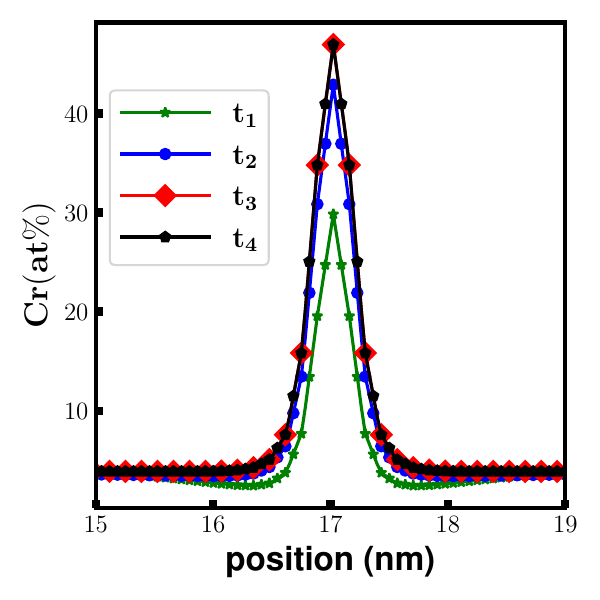}
\caption{Evolution of Cr concentration with time for Fe-5at\% Cr at 700K. The time $t_1$ is 200 sec, $t_2$ is 500 sec, $t_3$ is 2000 sec and $t_4$ is 2500 sec.}
\label{fig:cr_evo}
\end{figure}
Our phase field simulation begins with a bicrystal configuration featuring a flat GB. In Figure~\ref{fig:schematic}(a), we illustrate this bicrystal setup. Each grain possesses a distinct value of $\eta$, as depicted in Figure~\ref{fig:schematic}(b). Specifically, $\eta_1$ is 1 within grain 1 and zero within grain 2, while $\eta_2$ is 1 within grain 2 and zero within grain 1. Within the GB region, the values of $\eta$ change smoothly. As a result of the flat GB, it remains stationary. The primary motivation for selecting this stationary GB configuration is to study equilibrium GB segregation with respect to the bulk concentration and temperature and validate against established GB segregation theories, such as the McLean grain boundary segregation isotherm~\cite{mclean1958grain}.

\begin{figure}[ht]
\centering 
\includegraphics[width=0.70\linewidth]{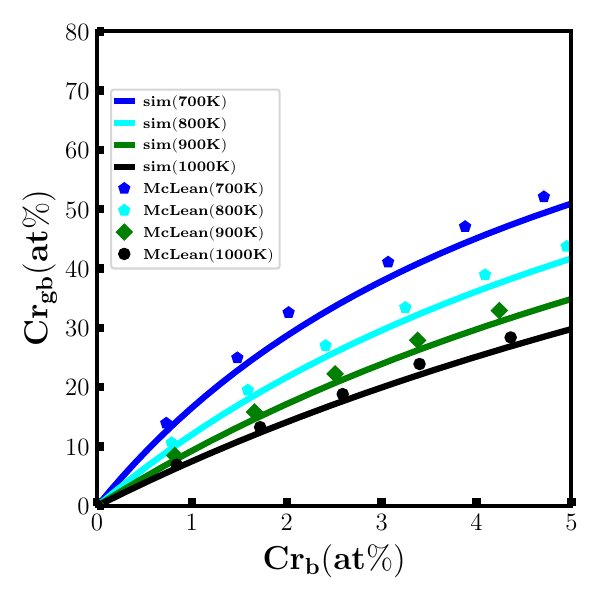}
\caption{Comparison of $Cr_{gb}$ vs. $Cr_{b}$ plot between simulation results and McLean isotherm [Equation~\ref{eq:McLean}]. $Cr_{b}$ refers to the equilibrium bulk concentration, and $Cr_{gb}$ refers to the equilibrium concentration at the centre of grain boundary.}
\label{fig:C_vs_T}
\end{figure}

We study GB segregation in bi-crystal over a range of initial Cr concentrations (1 at\% to 5 at\%) at four distinct temperatures (700K, 800K, 900K, and 1000K). Initially, at the beginning of the simulation, the Cr concentration remains uniform across the entire simulation domain. Over time, Cr atoms commence segregating at the GB, resulting in a decrease in bulk concentration. This segregation process continues until both bulk and GB concentrations reach equilibrium. 

Figure~\ref{fig:cr_evo} illustrates the change in Cr concentration with time for 700K for a material containing 5 at\% Cr. It becomes evident that as time progresses, Cr atoms from the bulk on both sides of the GB begin to accumulate at the GB. Atoms of Cr from neighboring grains on either side of the GB migrate toward it at an identical rate, resulting in their accumulation at the GB due to stationary GB. As a result, the flux on both sides of the GB becomes equal $(|J_f^-| = |J_f^+|)$, with the symbol $J_f$ representing Fickian flux. This atomic migration persists until the segregation of atoms at the GB reaches an equilibrium state with the bulk phase. It is evident from Figure~\ref{fig:cr_evo} that the peak segregation amount at the middle of the GB increases with increasing time (from $t_1$ to $t_2$). However, it reaches equilibrium at $t_3$; the Cr concentration profile remains unchanged after that. At $t_3$ and $t_4$, the concentration remains unchanged as it reaches the equilibrium at $t_3$. 



Using $E_{seg}$ values in the McLean isotherm \cite{mclean1958grain,KRAU202073}, we determine GB concentrations at specific temperatures for given bulk concentrations, assuming ideal mixing of solute and solvent atoms. The model is represented as:
\begin{equation}
C_{\text{GB}} = \frac{C_{0} \exp\left(-\frac{E_{\text{seg}}}{k_B T}\right)}{1 - C_{0} + C_{0} \exp\left(-\frac{E_{\text{seg}}}{k_{B} T}\right)}. 
\label{eq:McLean}
\end{equation}
Here $C_{0}$ is the equilibrium bulk concentration, $C_{GB}$ represents the equilibrium GB concentration, $k_{B}$ denotes the Boltzmann constant, and $T$ is the temperature. Figure~\ref{fig:C_vs_T} depicts the equilibrium GB Cr concentration ($Cr_{gb}$) for a given bulk Cr concentration ($Cr_{b}$) at four distinct temperatures (700K, 800K, 900K, and 1000K). The figure also compares our simulation outcomes against the McLean isotherm (Equation~\ref{eq:McLean}). We observe a noticeable trend: GB segregation increases as the bulk concentration rises. However, GB segregation reduces at elevated temperatures, consistent with established theoretical predictions~\cite{mclean1958grain}. The agreement between our simulation results and the McLean isotherm affirms our model's precision.


\subsection{Polycrystal system}
\begin{figure}[ht]
\centering 
\includegraphics[width=0.5\linewidth]{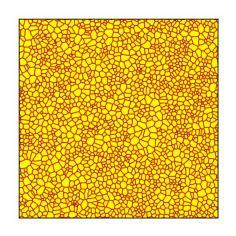}
\caption{Initial microstructure for all simulation having 2000 grains in a system size of $1600 \times 1600$ nm$^2$.}
\label{fig:init_micro}
\end{figure}

\begin{figure}[htpb]
\centering 
\includegraphics[width=0.99\linewidth]{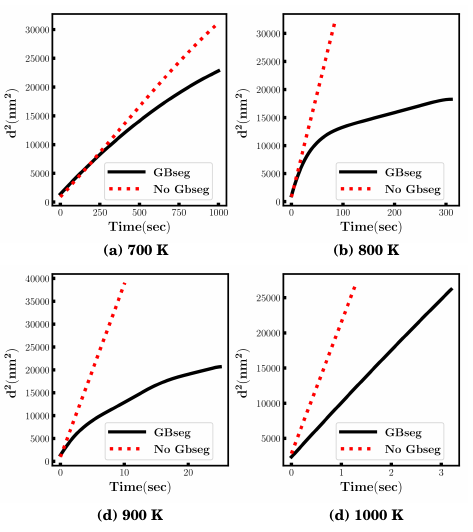}
\caption{Comparison of $d^2(nm^2)$ vs. $t(sec)$ plot; with Cr segregation (black solid line) and without Cr segregation (red dotted line).}
\label{fig:a_vs_t}
\end{figure}
We now focus on the nanostructured polycrystalline system and perform simulations at four distinct temperatures. We examine the influence of Cr segregation on the grain growth dynamics within the single phase $\alpha$-Fe, starting with an initial Cr concentration of 5 at\% across all simulations. We compare the results with and without GB segregation at each temperature, aiming to evaluate how GB segregation influences grain growth kinetics in nanocrystalline materials. In simulations without segregation, $E_{seg}$ is set to zero, while other parameters remain consistent for both sets of simulations. Initially, the microstructure contains 2000 grains in a $1600 \times 1600$ nm$^2$ system, resulting in an average grain diameter ($d_o$) of 35.7 nm. Figure~\ref{fig:init_micro}(a) shows the initial microstructure for all simulations. The kinetics of grain growth in materials science is often described by the following mathematical relationship known as the grain growth kinetics formula~\cite{WANG20211391}:
\begin{equation}          
           d^n=d^n_o+Kt. 
          \label{eq:growth_kinetics}
\end{equation}
Here, $d$ represents the average grain size, $d_o$ is the initial grain size at time t=0, $K$ is a temperature-dependent rate constant, and $n$ is a parameter that characterizes the grain growth mechanism. The value of $n$ depends on the specific type of grain growth occurring, with typical values being 2 for normal grain growth~\cite{WANG20211391}. Figures~\ref{fig:a_vs_t}(a-d) illustrates the comparison of $d^2$ vs. $t$ between the cases with GB segregation and without GB segregation for all four temperatures. The red dashed line represents the case without segregation, while the black solid line signifies a situation with Cr segregation. At all temperatures, the $d^2$ vs. $t$ plot follows a linear pattern for the case without GB segregation. Conversely, when GB segregation is present, the $d^2$ vs. $t$ plot deviates from the linear trend.

\begin{figure}[ht]
\centering 
\includegraphics[width=0.95\linewidth]{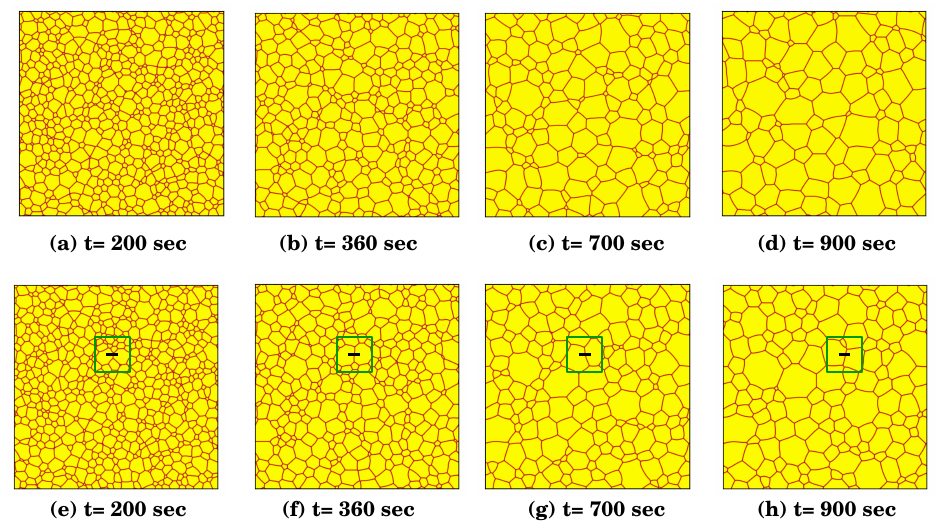}
\caption{Evolution of polycrystalline microstructure at 700K without Cr segregation at (a) 200 sec, (b) 360 sec, (c) 700 sec, and (d) 900 sec. Evolution of polycrystalline microstructure at 700K with Cr segregation at (e) 200 sec, (f) 360 sec, (g) 700 sec, and (h) 900 sec.   }
\label{fig:700K_micro}
\end{figure}

\begin{figure}[ht]
\centering 
\includegraphics[width=0.95\linewidth]{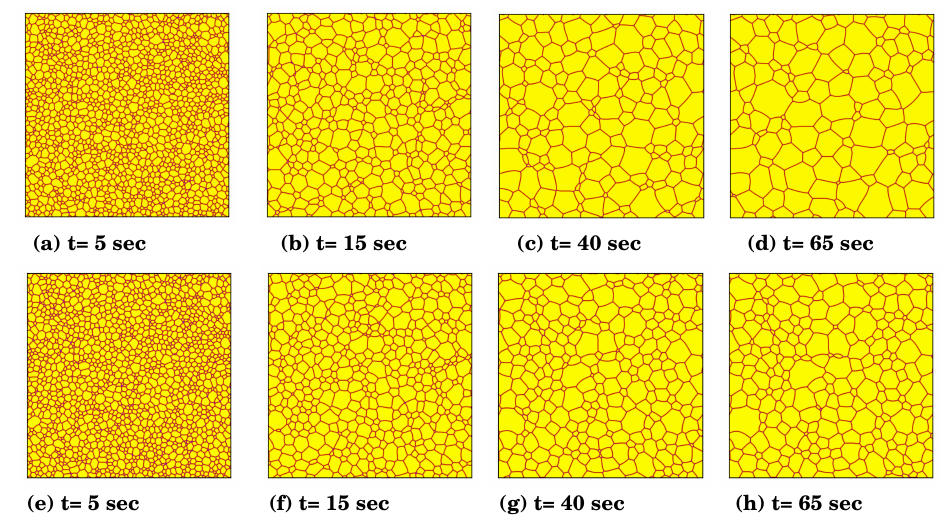}
\caption{Evolution of polycrystalline microstructure at 800K without Cr segregation at (a) 5 sec, (b) 15 sec, (c) 40 sec, and (d) 65 sec. Evolution of polycrystalline microstructure at 800K with Cr segregation at (e) 5 sec, (f) 15 sec, (g) 40 sec, and (h) 65 sec. }
\label{fig:800K_micro}
\end{figure}

\begin{figure}[ht]
\centering 
\includegraphics[width=0.95\linewidth]{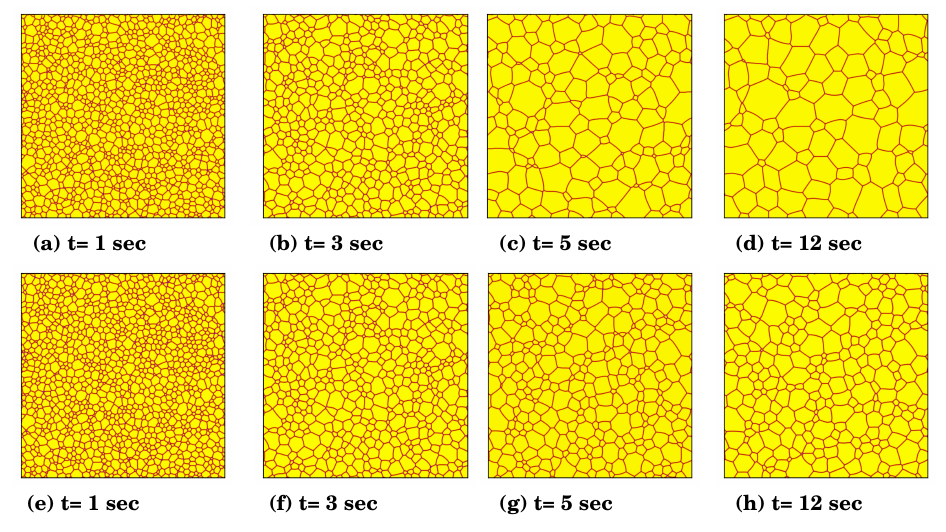}
\caption{Evolution of polycrystalline microstructure at 900K without Cr segregation at (a) 1 sec, (b) 3 sec, (c) 5 sec, and (d) 12 sec. Evolution of polycrystalline microstructure at 900K with Cr segregation at (e) 1 sec, (f) 3 sec, (g) 5 sec, and (h) 12 sec. }
\label{fig:900K_micro}
\end{figure}

\begin{figure}[ht]
\centering 
\includegraphics[width=0.95\linewidth]{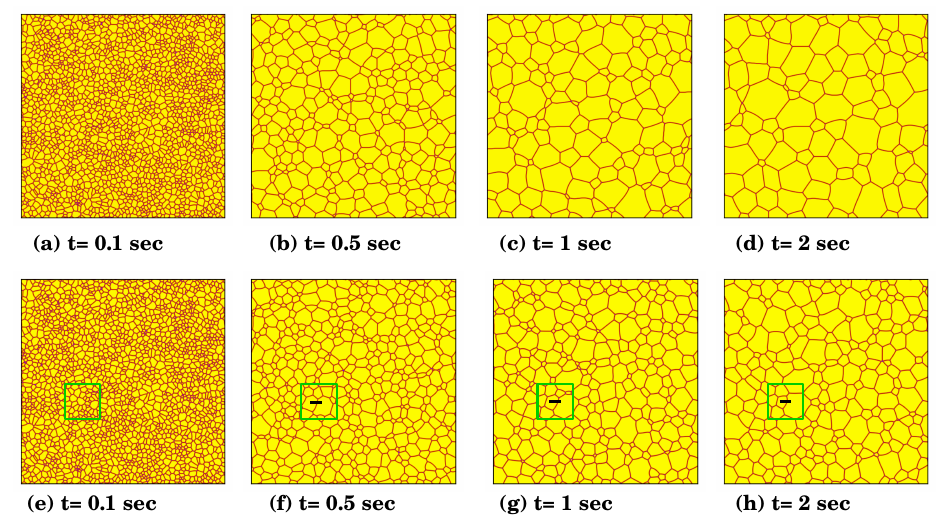}
\caption{Evolution of polycrystalline microstructure at 1000K without Cr segregation at (a) 0.1 sec, (b) 0.5 sec, (c) 1 sec, and (d) 2 sec. Evolution of polycrystalline microstructure at 1000K with Cr segregation at (e) 0.1 sec, (f) 0.5 sec, (g) 1 sec, and (h) 2 sec.}
\label{fig:1000K_micro}
\end{figure}

\begin{figure}[ht]
\centering 
\includegraphics[width=0.75\linewidth]{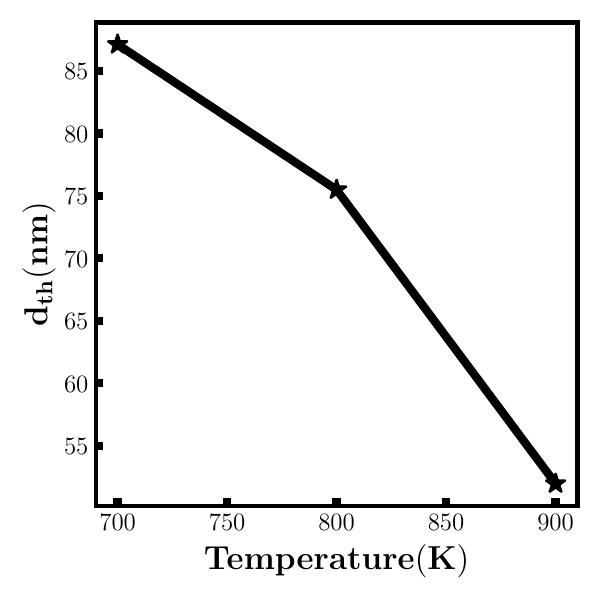}
\caption{ Variation of threshold size ($d_{th}$) with temperature (K).}
\label{fig:d_th}
\end{figure}

\begin{figure*}[ht]
\centering 
\includegraphics[width=0.90\linewidth]{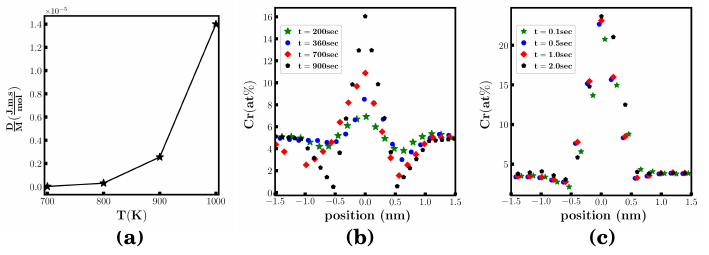}
\caption{ (a) Variation of $\frac{D}{M}$ with temperature (K). (b) Evolution of Cr composition at moving GB with time for 700K. This 1D composition plot is taken along the black solid line indicated in Figure~\ref{fig:700K_crop}. (c) Evolution of Cr composition at moving GB with time for 1000K. This 1D composition plot is taken along the black solid line indicated in Figure~\ref{fig:1000K_crop}.}
\label{fig:D_M}
\end{figure*}

\begin{figure}[ht]
\centering 
\includegraphics[width=0.95\linewidth]{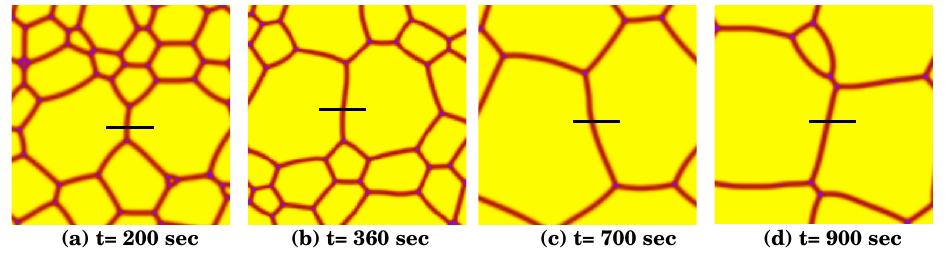}
\caption{Evolution of microstructure at 700K without Cr segregation at (a) 200 sec, (b) 360 sec, (c) 700 sec, and (d) 900 sec. These microstructures are the enlarged view of the region bounded by the green square in Figure~\ref{fig:700K_micro}(e-h).}
\label{fig:700K_crop}
\end{figure}

\begin{figure}[ht]
\centering 
\includegraphics[width=0.95\linewidth]{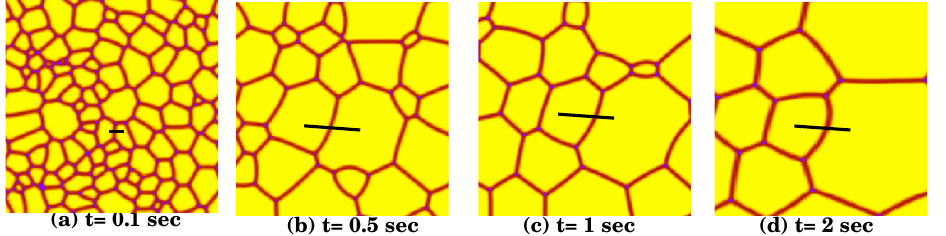}
\caption{Evolution of microstructure at 1000K with Cr segregation at (a) 0.1 sec, (b) 0.5 sec, (c) 1 sec, and (d) 2 sec. These microstructures are the enlarged view of the region bounded by the green square in Figure~\ref{fig:1000K_micro}(e-h).}
\label{fig:1000K_crop}
\end{figure}

Figures~\ref{fig:a_vs_t}(a-d) shows the $d^2$ vs. $t$ plot between with segregation and without segregation at different temperatures. At 700K, initially, both curves share the same slope up to a threshold grain size ($d_{th}$). However, after the threshold grain size ($d_{th}=87.12$ nm at time=360 sec), the curve representing the scenario with Cr segregation begins to deviate from its linear trajectory [Figure~\ref{fig:a_vs_t}a]. Corresponding microstructures at different times (t=200 sec, 360 sec, 700 sec, and 900 sec) are illustrated in Figures~\ref{fig:700K_micro}(a-d) for the case without GB segregation and in Figures~\ref{fig:700K_micro}(e-h) for the case with GB segregation. At 200 sec, both cases exhibit a similar grain growth rate, which is maintained up to 360 sec. However, at 700 and 900 sec, the microstructure with GB segregation contains more grains.

At 800K, the deviation of $d^2$ vs. $t$ curve for the GB segregation case starts around the threshold grain size of 75.49 nm at 15 sec [Figure~\ref{fig:a_vs_t}b]. Corresponding microstructures at different times (t= 5 sec, 15 sec, 40 sec, and 65 sec) are illustrated in Figures~\ref{fig:800K_micro}(a-d) for the case without GB segregation and in Figures~\ref{fig:800K_micro}(e-h) for the case with GB segregation. Up to time= 15 sec, both cases maintain a similar grain growth rate. However, at 40 and 65 sec, the microstructure with Cr segregation at the GB contains more grains, suggesting a significant slowdown of grain growth due to a higher solute drag force.

Similar to 700 and 800 K, a qualitatively similar trend is observed at 900 K. The $d^2$ vs. $t$ curve with GB segregation deviates from the linear trend beyond t= 3 sec and threshold grain size around 51.96 nm [Figure~\ref{fig:a_vs_t}c]. Corresponding microstructures at different times (t= 1 sec, 3 sec, 5 sec, and 12 sec) are illustrated in Figures~\ref{fig:900K_micro}(a-d) in the absence of GB segregation and Figures~\ref{fig:900K_micro}(e-h) in the presence of GB segregation. At 1 sec, both scenarios exhibit a similar grain growth rate, which is maintained up to 3 sec. However, a significantly slower grain growth, due to a higher solute drag force because of Cr segregation at GB, results in more grains at 5 sec and 12 sec.

At 1000 K, $d^2$ vs. $t$ curve for both with GB segregation and without segregation are linear but different slope [Figure~\ref{fig:a_vs_t}d]. For with GB segregation, the curve has lower slope. Corresponding microstructures at different times (t= 0.1 sec, 0.5 sec, 1 sec, and 2 sec) are illustrated in Figures~\ref{fig:1000K_micro}(a-d) in the absence of GB and Figures~\ref{fig:1000K_micro}(e-h) in the presence of GB segregation. At 0.5 sec and beyond, the microstructures with GB segregation contain more grains due to slower grain growth. The grain size distributions at 700K, 800K, 900K and 1000K are provided in the supplementary materials.


\textcolor{black}{Velocity of GB during grain growth in the presence of GB segregation is given as~\cite{HEO20117800};
\begin{equation}          
           V_{gb}=M[F_{gb}-P_{drag}], 
          \label{eq:v}
\end{equation}
here, $V_{gb}$ is the GB velocity, $P_{drag}$ is the drag force due to GB segregation, and $F_{gb}$ is curvature induced driving force for grain growth. For polycrystalline structure, $F_{gb}$ is inversely proportional to the size of grain. As the grain size increases, $F_{gb}$ decreases; hence $V_{gb}$ also decreases. The GB mobility (M) is also an important factor for determining the GB velocity. As the M value increases, the $V_{gb}$ also increases. $P_{drag}$ force in a moving GB depends on the amount of segregation occurring at the GB, which is also related to the diffusivity (D) of the solute atoms. As the D value increases, the amount of segregation at GB increases, resulting in a higher drag force~\cite{HEO20117800}. Therefore, with increasing temperature, both M and D values rise. Consequently, the increase in M value leads to higher $V_{gb}$, thus lower segregation, whereas an increase in D value leads to greater GB segregation. Consequently, with rising temperature, the extent of segregation in a moving GB can be influenced by both the values of D and M. }

\textcolor{black}{In the initial phase, the grain sizes are extremely small, resulting in significant grain growth propelled by a robust driving force ($F_{gb}$). During this stage, the velocity of GB ($V_{gb}$) is high, leading to minimal chromium (Cr) segregation at the GB, as higher GB velocity correlates with lower segregation~\cite{CAHN1962789, HEO20117800}. As time progresses, grain growth occur, causing a decline in the driving force for grain growth and reduction in GB velocity. Consequently, the amount of solute segregation at GB increases over time. As a result, as grain size increases, segregation at the moving GB also increases. This leads to a threshold grain size ($d_{th}$) beyond which a substantial amount of segregation occurs at GB, resulting in a significant drag force on the moving GB. Below this threshold size, the plot of $d^2$ vs. $t$ follows a linear trend same as case for without segregation. Beyond this point, however, the presence of segregation leads to a slower grain growth rate. Hence, $d^2$ vs. $t$ plot starts to deviate from its linear trend.}

\textcolor{black}{Figure~\ref{fig:d_th} shows the threshold size vs. temperature plot for 700K to 900K. It is clearly visible that with increasing temperature threshold grain size decreases. 
At 1000 K, $d^2$ vs. $t$ curve for both with and without segregation are linear but different slope [Figure~\ref{fig:a_vs_t}d]. 
To understand this grain growth behaviour due to Cr segregation at the GB, we focus on the temperature dependence of GB mobility ($M$) and Cr diffusivity ($D$). As shown in Figure~\ref{fig:D_M}a, the $\frac{D}{M}$ ratio increases as the temperature rises. Based on the plot, one can conclude that the increase of Cr diffusivity is more significant than the enhancement of GB mobility with increasing temperature. Thus, the Cr atoms can diffuse faster and segregate at the moving GB more easilty at higher temperatures. As a result, at 700 K, because of the low diffusivity of Cr, by the time sufficient GB segregation happens for the solute drag to be effective, the grains have grown to $d_{th}\sim 87$ nm at time $t= 360$ sec. After that, more Cr segregation occurs at the GB (due to grain growth), and solute drag effectively slows the grain growth process. Figure~\ref{fig:D_M}(b) further corroborates this, showing the increase of Cr concentration in moving GB at 700 K.
Note that concentration plots for chromium are obtained along a GB denoted by the small black lines in surrounded by green box. (shown in Figures~\ref{fig:700K_micro}(e-h)). Enlarge view of these microstructure (region bounded by the green lines) are shown in Figure~\ref{fig:700K_crop}. }

\textcolor{black}{Because of the enhanced diffusivity of Cr at higher temperatures, sufficient GB segregation for effective solute drag occurs earlier, such that the $d_{th}\sim$75 nm at 800 K and further reduces to $\sim$52 nm at 900 K. Figure~\ref{fig:d_th} shows how $d_{th}$ decreases as a function of temperature. The trend is almost inverse to the $\frac{D}{M}$ vs. temperature plot. Such a trend further proves that for Fe-5at\% Cr, $\frac{D}{M}$ plays a significant role in grain growth kinetics with Cr segregation. As the temperature increases, diffusivity increases more than GB's intrinsic mobility, resulting in sufficient drag force on moving GB. After sufficient Cr segregation at the GB, grain growth slows down due to the solute drag effect, and $d^2$ vs. $t$ deviates from the initial linear trend in all the cases.}

\textcolor{black}{Grain growth at 1000 K presents an interesting scenario, as the effect of segregation is manifested right from the beginning [Figure~\ref{fig:a_vs_t}d]. At 1000K, the diffusivity of Cr is so high that substantial segregation occurs very early. Notably, Cr-segregation reaches its maximum value [equilibrium value, determined by the McLean isotherm (Figure~\ref{fig:C_vs_T})] at a very early stage, and after that, the segregation level remains constant. It is evident from the 1D Cr concentration plot taken at GB at different times [Figure~\ref{fig:D_M}(c)]. 
Note that concentration plots for chromium are obtained along a GB denoted by the small black lines in surrounded by green box. (shown in Figures~\ref{fig:1000K_micro}(e-h)). Enlarged view of these microstructure (region bounded by the green lines) are shown in Figures~\ref{fig:1000K_crop}. Since the Cr-segregation level remains the same from the very early stage, the impact of solute drag remains similar throughout the grain growth process, resulting in a linear relationship with time. As a result, the $d^2$ vs. $t$ plot is also linear with Cr- segregation at the GB, with a lower slope (compared to no GB segregation) because of the solute drag effect. On the contrary, at a lower temperature, say 700 K, because of the lower $\frac{D}{M}$ ratio, the Cr-segregation at the grain boundary keeps increasing with time [Figure~\ref{fig:D_M}(b)] and is yet to reach its maximum value [equilibrium value, determined by the McLean isotherm (Figure~\ref{fig:C_vs_T})], even at t= 900 sec. As a result, the solute drag effect keeps increasing with time, causing the $d^2$ vs. $t$ curve to flatten gradually [Figure~\ref{fig:a_vs_t}] at 700 K.} 

\section{Discussion}

\textcolor{black}{Based on our previous results, it is evident that grain growth kinetics deviate from a linear trend when GB segregation occurs. This deviation is influenced by the value of the $\frac{D}{M}$ ratio. 
To further justify our findings, we developed an analytical formula to predict the evolution of grain size over time in the presence of GB segregation by solving Equation~\ref{eq:v} ($V_{gb}=M(F_{gb}-P_{drag})$).
In the case of a polycrystalline system, $F_{gb}=\frac{2\sigma_{gb}}{r}$ (because of curvature-driven GB migration).
Here, $r$ is the average grain size.
During grain growth, the curvature-driven driving force and the segregation-induced drag force ($P_{drag}$) simultaneously act on a moving grain boundary.
In a seminal work, Cahn formulated the drag force on a moving GB, which is expressed as~\cite{CAHN1961795};}

\begin{equation}          
           P_{drag}=\frac{\alpha V_{gb} C_o}{1+{\beta}^2 V_{gb}^2} 
          \label{eq:drag}
\end{equation}
\textcolor{black}{here, $V_{gb}$ is the GB migration velocity, and $C_o$ is the bulk concentration of solute. The formulation to calculate $\alpha$ (Equation~\ref{eq:alpha}) and $\beta^2$ (Equation~\ref{eq:beta}) is given as~\cite{CAHN1961795};}

\begin{equation}          
           \alpha=4N_vkT\int_{-
\infty}^{+\infty}\frac{sinh^2(E(x)/2kT)}{D(x)} dx 
          \label{eq:alpha}
\end{equation}

\begin{equation} 
           \frac{\alpha}{\beta^2}=\frac{N_v}{kT}\int_{-
\infty}^{+\infty}{\left\{\frac{dE(x)}{dx} \right\}}^2 D(x)  dx
        \label{eq:beta}
\end{equation}

\textcolor{black}{Here, $N_v$ is the molar volume. $E(x)$ is the GB solute interaction potential acting on the GB. By substituting the value of $P_{drag}$ from Equation~\ref{eq:drag} into Equation~\ref{eq:v}, we derive an analytical solution to determine the grain growth rate constant at any given time, expressed as follows;}

\begin{equation} 
\small
           K=  \frac{-\{16t(M\alpha C_o)-\beta^2K_o\}+\sqrt{\{16t(M\alpha C_o)-\beta^2K_o\}^2-64\beta^2K_ot}}{-32K_ot},  
          \label{eq:K}
\end{equation}
here, $K_o$ is the grain growth rate constant for the same system without any GB segregation. A detailed derivation of Equation~\ref{eq:K} is provided in the Appendix:A section. Equation~\ref{eq:K} allows us to calculate the value of $K$ at any given time.

\begin{figure}[ht]
\centering 
\includegraphics[width=0.70\linewidth]{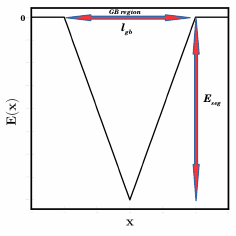}
\caption{Shape of GB solute interaction potential (E(x)). Here, $l_{gb}$ is the GB width and $E_{seg}$ is the segregation energy. }
\label{fig:E_schematic}
\end{figure}

One of the most challenging aspects of calculating
the value of $K$ using Equation~\ref{eq:K} is 
determining the values of $\alpha$ (using 
Equation~\ref{eq:alpha}) and ${\beta}^2$ (using
Equation~\ref{eq:beta}) due to the $E(x)$ function.
This $E(x)$ describes the solute-GB interaction potential function as described by Cahn~\cite{CAHN1961795}. 
Figure~\ref{fig:E_schematic} shows the schematic representation of the $E(x)$ function. The value of $E(x)$ is zero within the bulk, and its value is negative within the GB. Again, the most negative value of $E(x)$ is at the centre of the GB and from the centre point of GB, the value of $E(x)$ decreases (as shown in Figure~\ref{fig:E_schematic}).
Here, we introduce a new function to represent $E(x)$, keeping all the material information within itself. In this context, the materials information are GB width ($l_{gb}$) and the segregation energy ($E_{seg}$) value. In this case, we propose $E(x)$ as a function of $l_{gb}$ and $E_{seg}$, which is given as;

\begin{equation}          
           E(x)= 0.5E_{seg}\left\{|(1-|\frac{x}{0.5l_{gb}}|)| + (1-|\frac{x}{0.5l_{gb}}|)\right\}. 
          \label{eq:d_E}
\end{equation}

\textcolor{black}{This $E(x)$ function has the same shape as described by Figure~\ref{fig:E_schematic}. 
By substituting this $E(x)$ (as given by Equation~\ref{eq:d_E}) function into Equation~\ref{eq:alpha} and Equation~\ref{eq:beta} we are able to calculate the value of $\alpha$ and ${\beta}^2$; hence, the value of $K$ as function of time (t) using Equation~\ref{eq:K}. After the determination of $K$, we can calculate the evolution of grain size (by substituting $K$ in Equation~\ref{eq:growth_kinetics}) using the following;}

\begin{equation} 
    \small
   d^2= d^2_o + \left\{ \frac{-\{16t(M\alpha C_o)-\beta^2K_o\}+\sqrt{\{16t(M\alpha C_o)-\beta^2K_o\}^2-64\beta^2K_ot}}{-32K_ot}\right\}t, 
  \label{eq:d2_t}
\end{equation}

\begin{figure}[ht]
\centering 
\includegraphics[width=0.70\linewidth]{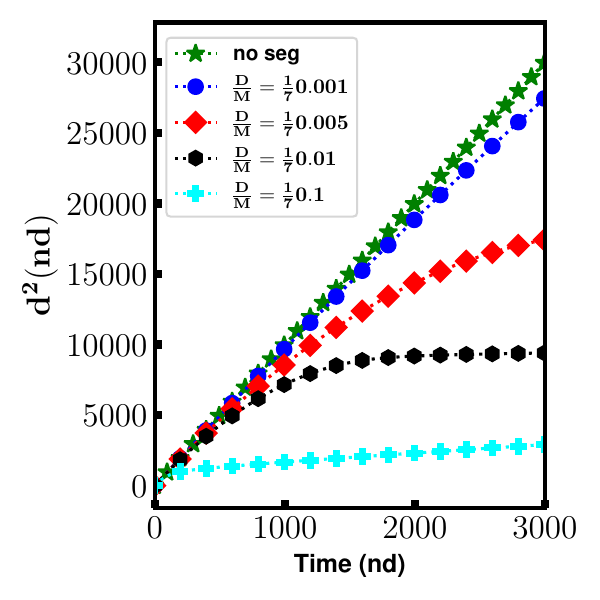}
\caption{$d^2(nd)$ vs. $Time(nd)$ plot for different $\frac{D}{M} $ values. Here, 'no seg' represents the case without segregation ($E_{seg}=0$), and the slope of this line is the same as $K_o=10.0$.  }
\label{fig:D_M_1}
\end{figure}

\textcolor{black}{Figure~\ref{fig:D_M_1} shows the effect of $\frac{D}{M}$ ratio in grain growth behaviour. Here, the evolution of grain size with time is calculated using
Equation~\ref{eq:d2_t}. In this case, we take a simple material system, where $M=7.0$, $C_o=0.05$, $K_o=10.0$, $l_{gb}=1.0$, $E_{seg}=-0.1$, $kT=0.04285$ and $N_v=1.0$. In this context, all the units are in non-dimensional (nd) form. After keeping all these values fixed, we choose different values of D. The D values are 0.001 ($\frac{D}{M}=\frac{1}{7}\times0.001$), 0.005 ($\frac{D}{M}=\frac{1}{7}\times0.005$), 0.01 ($\frac{D}{M}=\frac{1}{7}\times0.01$) and 0.1 ($\frac{D}{M}=\frac{1}{7}\times0.1$). It is observed that, without segregation ($E_{seg}=0$), $d^2(nd)$ vs $t(nd)$ follows the linear curve with a slope of $K_o$. As the D value increases, $d^2(nd)$ vs. $t(nd)$ plot initially follows the same slope (i.e. $K_o$) up to a threshold size ($d_{th}$), and after that, it starts to deviate with a smaller slope. Hence, it is visible that, depending on the $\frac{D}{M}$ value, grain growth kinetics can carry in the presence of GB segregation. It is also visible that, with increasing the $\frac{D}{M}$, the $d_{th}$ size decreases. This phenomenon is similar to our Fe-Cr polycrystal simulation as shown in Figure~\ref{fig:d_th}, which describes the reduction in $d_{th}$ with increasing temperature. As in the Fe-Cr system, with increasing the temperature, the $\frac{D}{M}$ value increases (shown in Figure~\ref{fig:D_M})(a), which leads to a reduction in $d_{th}$.
It is also visible from Figure~\ref{fig:D_M_1}, that at $\frac{D}{M}=\frac{1}{7}0.1$ (highest value of $\frac{D}{M}$), the $d_{th}$ value is almost same as initial grain size, which means the deviation of $d^2(nd)$ vs $t(nd)$ plot starts at the initial time, due to high value of D, which allows solute atoms to catch the moving GB, helping to achieve maximum possible segregation at initial stage of grain growth. This situation is similar to the case of grain growth at 1000K for the Fe-Cr system (shown in Figure~\ref{fig:a_vs_t}(d)).} 

\begin{figure}[ht]
\centering 
\includegraphics[width=0.99\linewidth]{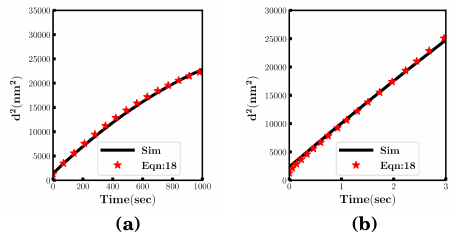}
\caption{ (a) Comparison of $d^2(nm^2)$ vs. $t(sec)$ plot (with the presence of Cr segregation) from simulation (black solid line) and prediction using Equation~\ref{eq:d2_t} (red star shaped dots) at 700K. (b) Comparison of $d^2(nm^2)$ vs. $t(sec)$ plot from simulation (black solid line) and prediction using Equation~\ref{eq:d2_t} (red star shaped dots) at 1000K.}
\label{fig:sim_calc}
\end{figure}

\textcolor{black}{We have compared the $d^2(nm^2)$ vs $t(sec)$ plot (with the presence of Cr segregation) between calculation using Equation~\ref{eq:d2_t} and our simulation. Figure~\ref{fig:sim_calc}(a) shows the comparison of $d^2(nm^2)$ vs $t(sec)$ plot in the presence of GB segregation between simulation results and calculation done using Equation~\ref{eq:d2_t} at 700K, while Figure~\ref{fig:sim_calc}b shows the same for 1000K. It is visible that our simulation results are well aligned with the prediction using Equation~\ref{eq:d2_t}. One important factor is that the derivation of Equation~\ref{eq:K} is based on the assumption that the GB energy does not change with the amount of GB segregation. In our phase-field simulation also, the change in GB energy with solute segregation is negligible. 
 As in our case, we are mainly focusing on the solute drag force and amount of GB segregation during grain growth; our model is capable of predicting grain growth kinetics in the presence of solute segregation.
}
\textcolor{black}{In our phase-field simulation, we assumed that all the GBs are ($\sum3(1 \bar{1} 1)[110]$). However, in real systems, different GBs are present within the same materials. To achieve more accurate grain growth kinetics in the presence of GB segregation, it is necessary to incorporate different GBs into our model. 
However, most of the GBs found in real systems have properties similar to the ($\sum3(1 \bar{1} 1)[110]$) GB we considered in our study. }
\textcolor{black}{For example, energies of naturally occurring high angle GBs which were estimated from ab-intio for $\sum3(2\bar{2} 1)[110]$ and $\sum 11(3\bar{3} 2)[110]$ GBs $\sigma_{gb}$ have values around 1.75$J/m^2$ and 1.45$J/m^2$, and their $E_{seg}$ were reported to be -0.21$eV$ and -0.29$eV$ \cite{MAI2022117902}. These values are close to the values we considered for our simulation (given in Table~\ref{Tab:fe_cr_1}). Hence, our model is suitable for studying grain growth kinetics in the presence of solute segregation for binary alloys having high-angle tilt GBs. }

\section{Conclusion}\label{sec5}

\begin{itemize}

\item  We use a multiscale modelling framework, including DFT calculations and phase-field simulations, to study grain growth in nanocrystalline single-phase $\alpha$-Fe alloy with Cr segregation. Initially, DFT calculations provided key parameters such as GB width, energy, and Cr segregation energy, which were then used in phase-field simulations.

\item The bi-crystal simulations featuring a stationary flat GB aimed to validate our model accuracy against established GB segregation theories. Over a temperature range of 700K to 1000K and Cr compositions spanning 1 at\% to 5 at\%, our findings showcased a gradual segregation of Cr atoms at the GB, leading to a discernible decline in bulk concentration until reaching equilibrium. A comparative analysis with the McLean equation highlighted a consistent trend: heightened bulk concentration correlated with increased GB segregation, while elevated temperatures corresponded to diminished GB segregation, aligning with theoretical expectations.

\item In the polycrystal system, our simulations at temperatures ranging from 700K to 1000K, maintaining a constant initial Cr concentration of 5 at\%, revealed the substantial impact of GB segregation on grain growth dynamics. The presence of GB segregation notably altered the linear relationship of $d^2$ vs. $t$, emphasizing the intricate interplay between segregation, solute drag force, and the rate of grain growth. 

\item We explored grain growth behaviour within nanocrystalline grain structures at varying temperatures. Notably, our investigation revealed that the nature of grain growth is influenced by the $\frac{D}{M}$ ratio. 

\item The threshold grain size ($d_{th}$) at which an effective drag force begins to influence grain boundaries is temperature-dependent. For the Fe-5at\% Cr system in a single-phase $\alpha$-Fe, the values of $d_{th}$ are 87.12 nm, 75.49 nm, and 51.96 nm at temperatures of 700K, 800K, and 900K, respectively. At 1000K, maximum segregation occurs at the initial grain size.

\item \textcolor{black}{We derive an analytical solution to predict the grain size evolution (in the presence of GB solute segregation) with time during grain growth. Our simulation results are well aligned with the analytical solution.}

These insights enhance our understanding of atomic-scale phenomena, offering implications for the strategic design and enhancement of materials with tailored mechanical and structural properties.

\end{itemize}

\section*{Acknowledgments}

RM and SB acknowledges financial support from SERB core research grant (CRG/2019/006961). The authors acknowledge National Supercomputing Mission (NSM) for providing computing resources of ``PARAM Sanganak'' at IIT Kanpur, which is implemented by C-DAC and supported by the Ministry of Electronics and Information Technology (MeitY) and Department of Science and Technology (DST), Government of India. Authors also thank ICME National Hub, IIT Kanpur, and computer center, IIT Kanpur, for providing HPC facility.

\section*{Data Availability Statement}
The data that support the findings of this study are available from the corresponding author upon reasonable request.

\section*{CRediT authorship contribution statement}
\textbf{Sandip Guin:} Conceptualization, Visualization, Methodology, Software, Investigation, Formal analysis, Validation, Data curation, Writing-Original Draft. \textbf{Albert Linda:} Visualization, Methodology, Software, Investigation, Formal analysis, Validation, Data curation, Writing-Original Draft. \textbf{Yu-Chie Lo:} Supervision, Project administration, Resources, Writing - review $\&$ editing. \textbf{Somnath Bhowmick:} Supervision, Project administration, Resources, Writing - review $\&$ editing, Funding acquisition. $\&$ editing. \textbf{Rajdip Mukherjee:} Supervision, Project administration, Resources, Writing - review $\&$ editing, Funding acquisition.

\appendix
\section{Grain growth rate constant derivation}
\label{appendix2}
The GB migration velocity in the presence of GB solute segregation is given as~\cite{HEO20117800};
\begin{equation}
    V_{gb} = M\left\{F_{gb}-P_{drag}\right\},
    \label{eq:A1}
\end{equation}
here, $F_{gb}$ is the driving force for GB migration. During the grain growth in  polycrystalline system, $F_{gb}=\frac{2\sigma_{gb}}{r}$. Here, r is the average grain size. So, substituting the value of $F_{gb}$ in Equation~\ref{eq:A1} we get; 

\begin{equation}
    V_{gb} = M\left\{\frac{2\sigma_{gb}}{r}-P_{drag}\right\},
    \label{eq:A2}
\end{equation}
here, the $P_{drag}$ is given as~\cite{CAHN1961795}; 

\begin{equation}
    P_{drag} = \frac{\alpha V_{gb} C_o}{1+{\beta}^2 V_{gb}^2},
    \label{eq:A3}
\end{equation}
in this $P_{drag}$ equation, $\alpha$ and $\beta^2$ are materials dependent parameter which is given by Equation~\ref{eq:alpha} and Equation~\ref{eq:beta}. Now after substituting $P_{drag}$ in Equation~\ref{eq:A2}, we get;

\begin{equation}
    V_{gb}\left\{ 1 + \frac{M \alpha c_o}{1+{\beta}^2 V_{gb}^2} \right\}=\frac{2M\sigma_{gb}}{r}.
    \label{eq:A4}
\end{equation}

During grain growth, the evolution of grain size with time is given as follows;
\begin{equation}
    d^2=d_o^2+Kt,
    \label{eq:A5}
\end{equation}
here $d$ is average grain size at time $t$, while $d_o$ is the initial grain size. $K$ is the grain growth rate constant, which determines the slope of $d^2$ vs. $t$ plot.
Replacing $r=\frac{d}{2}$ in Equation~\ref{eq:A5} and derivative wrt. t we get;

\begin{equation}
    8r\frac{\partial r}{\partial t}=K,
    \label{eq:A6}
\end{equation}
as $\frac{\partial r}{\partial t}=V_{gb}$, we can write as follows;

\begin{equation}
    V_{gb}=\frac{K}{8r}.
    \label{eq:A7}
\end{equation}

Substituting $V_{gb}$ in Equation~\ref{eq:A4}, we get as follows;

\begin{equation}
    \frac{K}{8r}\left\{ 1+\frac{M \alpha c_o}{1+{\beta}^2 V_{gb}^2}\right\}= \frac{2M\sigma_{gb}}{r},
    \label{eq:A8}
\end{equation}

\begin{equation}
    {K}\left\{ 1+\frac{M\alpha c_o}{1+{\beta}^2 V_{gb}^2}\right\}= 16M\sigma_{gb}.
    \label{eq:A9}
\end{equation}

Let's assume $16M\sigma_{gb}=K_o$, then we can rewrite as follows;

\begin{equation}
    {K}\left\{ 1+\frac{M_o \alpha c_o}{1+{\beta}^2 V_{gb}^2}\right\}= K_o,
    \label{eq:A10}
\end{equation}

\begin{equation}
    \frac{M \alpha c_o}{1+{\beta}^2 V_{gb}^2}= \frac{K_o}{K}-1.
    \label{eq:A11}
\end{equation}

From Equation~\ref{eq:A7}, $V_{gb}=\frac{K}{4d}$ (as $d=2r$). Substituting $V_{gb}$ in Equation~\ref{eq:A11} we get as follows;

\begin{equation}
    16d^2(M\alpha C_o+1)=\frac{K_o}{K}16d^2+{\beta}^2 K_o K - \beta^2K^2. 
    \label{eq:A12}
\end{equation}

During the grain growth process, if the initial grain size is very small ($d_o=0$), then we can write from Equation~\ref{eq:A5} that $d^2=Kt$. Now substituting this $d^2$ in Equation~\ref{eq:A12} we get as follows;

\begin{equation}
    {\beta}^2K^2+K\left\{ 16t(M\alpha C_o+1) - {\beta}^2K_o  \right\} -16K_ot=0.
    \label{eq:A13}
\end{equation}

The solution of the Equation~\ref{eq:A13} is as follows; 

\begin{equation}     
           K=  \frac{-\{16t(M\alpha C_o)-\beta^2K_o\}+\sqrt{\{16t(M\alpha C_o)-\beta^2K_o\}^2-64\beta^2K_ot}}{-32K_ot}.  
          \label{eq:A14}
\end{equation}

In absence of GB segregation ($P_{drag}=0$), Equation~\ref{eq:A2} becomes;

\begin{equation}
    V_{gb} = M\left\{\frac{2\sigma_{gb}}{r}\right\},
    \label{eq:A15}
\end{equation}

\begin{equation}
    \frac{K}{8r} = M\left\{\frac{2\sigma_{gb}}{r}\right\},
    \label{eq:A16}
\end{equation}

\begin{equation}
    K = 16M\sigma_{gb},
\end{equation}
so without GB segregation, K is a constant, which is given as $16M\sigma_{gb}$. So, from the slope of $d^2$ vs. $t$ plot in the absence of GB segregation, we can also calculate the value of $16M\sigma_{gb}$. This value is the same as $K_o$ in Equation~\ref{eq:A14}.

\bibliographystyle{unsrt}

\bibliography{mybibfile}

\begin{thebibliography}{10}

\bibitem{WU2023106083}
Zaoming Wu, Qiang Li, Xiaofeng Yang, Xiaoqiu Ye, Jipeng Zhu, and Jiliang Wu.
\newblock Microstructures and properties of nanocrystalline {W}-based alloys produced by resistance sintering under ultrahigh pressure.
\newblock {\em Materials Today Communications}, 35:106083, 2023.

\bibitem{GLEITER1989223}
H.~Gleiter.
\newblock Nanocrystalline materials.
\newblock {\em Progress in Materials Science}, 33(4):223--315, 1989.

\bibitem{BU2023171238}
Yifan Bu, Xiuzhen Zhang, and Dengshan Zhou.
\newblock Unraveling the strain-dependent hall-petch slope in low-to-high mg content {Al}-{Mg} alloys.
\newblock {\em Journal of Alloys and Compounds}, 963:171238, 2023.

\bibitem{HAMDI2023170826}
Hedayat Hamdi, Hamid~Reza Abedi, and Yong Zhang.
\newblock A review study on thermal stability of high entropy alloys: Normal/abnormal resistance of grain growth.
\newblock {\em Journal of Alloys and Compounds}, 960:170826, 2023.

\bibitem{DONG2023168515}
Jianxin Dong, Hongxing Wu, Ying Chen, Pengfei Li, Fan Zhang, Yunjie Wu, Ke~Hua, and Haifeng Wang.
\newblock Revealing the nano-grained microstructure and mechanical properties of electrochemical boronized alcocrfeni2.1 eutectic high entropy alloy.
\newblock {\em Journal of Alloys and Compounds}, 938:168515, 2023.

\bibitem{PENG2023168008}
Peng Peng, Jia She, Qingshan Yang, Shuai Long, Aitao Tang, Jianyue Zhang, Qingwei Dai, and Fusheng Pan.
\newblock Bimodal grained {Mg}–0.5{Gd}–x{Mn} alloys with high strength and low-cost fabricated by low-temperature extrusion.
\newblock {\em Journal of Alloys and Compounds}, 935:168008, 2023.

\bibitem{JINLONG2023167791}
Lv~Jinlong, Zhou Zhiping, Tong Liu, and Zhang Shuye.
\newblock Effects of heterogeneous ultrafine grain and strain rate on mechanical properties of cocrni medium entropy alloy.
\newblock {\em Journal of Alloys and Compounds}, 934:167791, 2023.

\bibitem{QIN2010S439}
Liyuan Qin, Jianshe Lian, and Qing Jiang.
\newblock Enhanced ductility of high-strength electrodeposited nanocrystalline {Ni}–{Co} alloy with fine grain size.
\newblock {\em Journal of Alloys and Compounds}, 504:S439--S442, 2010.
\newblock 16th International Symposium on Metastable, Amorphous and Nanostructured Materials.

\bibitem{DRIVER2004819}
J.H Driver.
\newblock Stability of nanostructured metals and alloys.
\newblock {\em Scripta Materialia}, 51(8):819--823, 2004.
\newblock Viewpoint set no. 35. Metals and alloys with a structural scale from the micrometer to the atomic dimensions.

\bibitem{GERTSMAN1994577}
V.Y. Gertsman and R.~Birringer.
\newblock On the room-temperature grain growth in nanocrystalline copper.
\newblock {\em Scripta Metallurgica et Materialia}, 30(5):577--581, 1994.

\bibitem{AMES20084255}
Markus Ames, Jürgen Markmann, Rudolf Karos, Andreas Michels, Andreas Tschöpe, and Rainer Birringer.
\newblock Unraveling the nature of room temperature grain growth in nanocrystalline materials.
\newblock {\em Acta Materialia}, 56(16):4255--4266, 2008.

\bibitem{GLEITER20001}
H.~Gleiter.
\newblock Nanostructured materials: basic concepts and microstructure.
\newblock {\em Acta Materialia}, 48(1):1--29, 2000.

\bibitem{D0NR07180C}
Christopher~M. Barr, Stephen~M. Foiles, Malek Alkayyali, Yasir Mahmood, Patrick~M. Price, David~P. Adams, Brad~L. Boyce, Fadi Abdeljawad, and Khalid Hattar.
\newblock The role of grain boundary character in solute segregation and thermal stability of nanocrystalline {Pt–Au}.
\newblock {\em Nanoscale}, 13:3552--3563, 2021.

\bibitem{MILLETT20072329}
Paul~C. Millett, R.~Panneer Selvam, and Ashok Saxena.
\newblock Stabilizing nanocrystalline materials with dopants.
\newblock {\em Acta Materialia}, 55(7):2329--2336, 2007.

\bibitem{WEISSMULLER1993261}
J.~Weissmüller.
\newblock Alloy effects in nanostructures.
\newblock {\em Nanostructured Materials}, 3(1):261--272, 1993.
\newblock Proceedings of the First International Conference on Nanostructured Materials.

\bibitem{LEJOEK2016}
P.~Lejoek and S.~Hofmann.
\newblock Anisotropy and quantitative prediction of grain boundary segregation.
\newblock In {\em Reference Module in Materials Science and Materials Engineering}. Elsevier, 2016.

\bibitem{YE2021102808}
Wenye Ye, Jake Hohl, Mano Misra, Yiliang Liao, and Leslie~T. Mushongera.
\newblock Grain boundary relaxation in doped nano-grained aluminum.
\newblock {\em Materials Today Communications}, 29:102808, 2021.

\bibitem{Lejcek201783}
Pavel Lejček, Mojmír Šob, and Václav Paidar.
\newblock Interfacial segregation and grain boundary embrittlement: An overview and critical assessment of experimental data and calculated results.
\newblock {\em Progress in Materials Science}, 87:83 – 139, 2017.

\bibitem{Raabe2014253}
D.~Raabe, M.~Herbig, S.~Sandlöbes, Y.~Li, D.~Tytko, M.~Kuzmina, D.~Ponge, and P.-P. Choi.
\newblock Grain boundary segregation engineering in metallic alloys: A pathway to the design of interfaces.
\newblock {\em Current Opinion in Solid State and Materials Science}, 18(4):253 – 261, 2014.

\bibitem{KIRCHHEIM2002413}
Reiner Kirchheim.
\newblock Grain coarsening inhibited by solute segregation.
\newblock {\em Acta Materialia}, 50(2):413--419, 2002.

\bibitem{HILLERT1976731}
Mats Hillert and Bo~Sundman.
\newblock A treatment of the solute drag on moving grain boundaries and phase interfaces in binary alloys.
\newblock {\em Acta Metallurgica}, 24(8):731--743, 1976.

\bibitem{CAHN1962789}
John~W Cahn.
\newblock The impurity-drag effect in grain boundary motion.
\newblock {\em Acta Metallurgica}, 10(9):789--798, 1962.

\bibitem{CANTWELL20141}
Patrick~R. Cantwell, Ming Tang, Shen~J. Dillon, Jian Luo, Gregory~S. Rohrer, and Martin~P. Harmer.
\newblock Grain boundary complexions.
\newblock {\em Acta Materialia}, 62:1--48, 2014.

\bibitem{WANG2021158326}
Zhenyu Wang, Zheng Chen, Yu~Fan, Jiachun Shi, Yuyu Liu, Xiao Shi, and Jie Xu.
\newblock Thermal stability of the multicomponent nanocrystalline ni–{ZrNbMoTa} alloy.
\newblock {\em Journal of Alloys and Compounds}, 862:158326, 2021.

\bibitem{DARLING2008530}
Kris~A. Darling, Ryan~N. Chan, Patrick~Z. Wong, Jonathan~E. Semones, Ronald~O. Scattergood, and Carl~C. Koch.
\newblock Grain-size stabilization in nanocrystalline {FeZr} alloys.
\newblock {\em Scripta Materialia}, 59(5):530--533, 2008.

\bibitem{HEO20117800}
Tae~Wook Heo, Saswata Bhattacharyya, and Long-Qing Chen.
\newblock A phase field study of strain energy effects on solute–grain boundary interactions.
\newblock {\em Acta Materialia}, 59(20):7800--7815, 2011.

\bibitem{WANG2023167717}
Lei Wang and Reza {Darvishi Kamachali}.
\newblock Calphad integrated grain boundary co-segregation design: Towards safe high-entropy alloys.
\newblock {\em Journal of Alloys and Compounds}, 933:167717, 2023.

\bibitem{MONDAL2014206}
R.A. Mondal, B.S. Murty, and V.R.K. Murthy.
\newblock Temperature and frequency dependent electrical properties of nicuzn ferrite with cuo-rich grain boundary segregation.
\newblock {\em Journal of Alloys and Compounds}, 595:206--212, 2014.

\bibitem{ITO2024107849}
Kazuma Ito.
\newblock Significant effect of magnetism on grain boundary segregation in $\gamma$-fe: A systematic comparison of grain boundary segregation in nonmagnetic and paramagnetic $\gamma$-fe by first-principles calculations.
\newblock {\em Materials Today Communications}, 38:107849, 2024.

\bibitem{MA2020101388}
Haibin Ma, Xinkai Ding, Libo Zhang, Yuanjun Sun, Tong Liu, Qisen Ren, and Yehong Liao.
\newblock Segregation of interstitial light elements at grain boundaries in molybdenum.
\newblock {\em Materials Today Communications}, 25:101388, 2020.

\bibitem{mclean1958grain}
Donald McLean and AJPT Maradudin.
\newblock Grain boundaries in metals, 1958.

\bibitem{KAUR2020109685}
Navjot Kaur, Chuang Deng, and Olanrewaju~A. Ojo.
\newblock Effect of solute segregation on diffusion induced grain boundary migration studied by molecular dynamics simulations.
\newblock {\em Computational Materials Science}, 179:109685, 2020.

\bibitem{GRONHAGEN2007955}
Klara Grönhagen and John Ågren.
\newblock Grain-boundary segregation and dynamic solute drag theory—a phase-field approach.
\newblock {\em Acta Materialia}, 55(3):955--960, 2007.

\bibitem{Zhao_2017}
Yu~Zhao, Shenhua Song, Hong Si, and Kai Wang.
\newblock Effect of grain size on grain boundary segregation thermodynamics of phosphorus in interstitial-free and 2.25{Cr-1Mo} steels.
\newblock {\em Metals}, 7(11):470, November 2017.

\bibitem{MAI2022117902}
Han~Lin Mai, Xiang-Yuan Cui, Daniel Scheiber, Lorenz Romaner, and Simon~P. Ringer.
\newblock The segregation of transition metals to iron grain boundaries and their effects on cohesion.
\newblock {\em Acta Materialia}, 231:117902, 2022.

\bibitem{HU2020109271}
Yong-Jie Hu, Yi~Wang, William~Y. Wang, Kristopher~A. Darling, Laszlo~J. Kecskes, and Zi-Kui Liu.
\newblock Solute effects on the $\sum3$ 111[11-0] tilt grain boundary in bcc {Fe}: Grain boundary segregation, stability, and embrittlement.
\newblock {\em Computational Materials Science}, 171:109271, 2020.

\bibitem{10.1063/1.4867400}
Hao Jin, Ilya Elfimov, and Matthias Militzer.
\newblock {Study of the interaction of solutes with $\sum$ 5 (013) tilt grain boundaries in iron using density-functional theory}.
\newblock {\em Journal of Applied Physics}, 115(9):093506, 03 2014.

\bibitem{ZHENG202040}
Hui Zheng, Xiang-Guo Li, Richard Tran, Chi Chen, Matthew Horton, Donald Winston, Kristin~Aslaug Persson, and Shyue~Ping Ong.
\newblock Grain boundary properties of elemental metals.
\newblock {\em Acta Materialia}, 186:40--49, 2020.

\bibitem{WANG2016279}
Shuai Wang, May~L. Martin, Ian~M. Robertson, and Petros Sofronis.
\newblock Effect of hydrogen environment on the separation of {Fe} grain boundaries.
\newblock {\em Acta Materialia}, 107:279--288, 2016.

\bibitem{RAZUMOVSKIY2018122}
V.I. Razumovskiy, S.V. Divinski, and L.~Romaner.
\newblock Solute segregation in {Cu}: Dft vs. experiment.
\newblock {\em Acta Materialia}, 147:122--132, 2018.

\bibitem{EBNER2021117354}
Anna~Sophie Ebner, Severin Jakob, Helmut Clemens, Reinhard Pippan, Verena Maier-Kiener, Shuang He, Werner Ecker, Daniel Scheiber, and Vsevolod~I. Razumovskiy.
\newblock Grain boundary segregation in {Ni}-base alloys: A combined atom probe tomography and first principles study.
\newblock {\em Acta Materialia}, 221:117354, 2021.

\bibitem{CHA20023817}
Pil-Ryung Cha, Seong~Gyoon Kim, Dong-Hee Yeon, and Jong-Kyu Yoon.
\newblock A phase field model for the solute drag on moving grain boundaries.
\newblock {\em Acta Materialia}, 50(15):3817--3829, 2002.

\bibitem{LiWangYang+2010+555+559}
Junjie Li, Jincheng Wang, and Gencang Yang.
\newblock Phase field simulation of grain growth with grain boundary segregation.
\newblock {\em International Journal of Materials Research}, 101(4):555--559, 2010.

\bibitem{KIM2016150}
Seong~Gyoon Kim, Jae~Sang Lee, and Byeong-Joo Lee.
\newblock Thermodynamic properties of phase-field models for grain boundary segregation.
\newblock {\em Acta Materialia}, 112:150--161, 2016.

\bibitem{VERMA2020155163}
Miral Verma and Rajdip Mukherjee.
\newblock Nanoparticle formation through dewetting of a solid-state thin film on a substrate: A phase-field study.
\newblock {\em Journal of Alloys and Compounds}, 835:155163, 2020.

\bibitem{GUO2023105811}
Can Guo, Ying Gao, Yu~teng Cui, Yu~ping Zhao, Chun jie Xu, Shang Sui, Xiang quan Wu, and Zhong ming Zhang.
\newblock Phase-field simulation of the spinodal decomposition process near moving grain boundaries.
\newblock {\em Materials Today Communications}, 35:105811, 2023.

\bibitem{LVOV2023106209}
Pavel~E. L’vov, Renat~T. Sibatov, and Vyacheslav~V. Svetukhin.
\newblock Anisotropic grain boundary diffusion in binary alloys: Phase-field approach.
\newblock {\em Materials Today Communications}, 35:106209, 2023.

\bibitem{WU2020100790}
Influencing factors of abnormal grain growth in mg alloy by phase field method.
\newblock {\em Materials Today Communications}, 22:100790, 2020.

\bibitem{KUNDIN2021110295}
Julia Kundin, Hedieh Farhandi, Kamatchi~Priya Ganesan, Renato~S.M. Almeida, Kamen Tushtev, and Kurosch Rezwan.
\newblock Phase-field modeling of grain growth in presence of grain boundary diffusion and segregation in ceramic matrix mini-composites.
\newblock {\em Computational Materials Science}, 190:110295, 2021.

\bibitem{VESTA}
Koichi Momma and Fujio Izumi.
\newblock {{\it VESTA3} for three-dimensional visualization of crystal, volumetric and morphology data}.
\newblock {\em Journal of Applied Crystallography}, 44(6):1272--1276, Dec 2011.

\bibitem{CHENG201892}
Jianli Cheng, Jian Luo, and Kesong Yang.
\newblock Aimsgb: An algorithm and open-source python library to generate periodic grain boundary structures.
\newblock {\em Computational Materials Science}, 155:92--103, 2018.

\bibitem{lejcek2010grain}
Pavel Lejcek.
\newblock {\em Grain boundary segregation in metals}, volume 136.
\newblock Springer Science \& Business Media, 2010.

\bibitem{CAHN1961795}
John~W Cahn.
\newblock On spinodal decomposition.
\newblock {\em Acta Metallurgica}, 9(9):795--801, 1961.

\bibitem{ALLEN19791085}
Samuel~M. Allen and John~W. Cahn.
\newblock A microscopic theory for antiphase boundary motion and its application to antiphase domain coarsening.
\newblock {\em Acta Metallurgica}, 27(6):1085--1095, 1979.

\bibitem{fazeli2005application}
F~Fazeli and M~Militzer.
\newblock Application of solute drag theory to model ferrite formation in multiphase steels.
\newblock {\em Metallurgical and Materials Transactions A}, 36:1395--1405, 2005.

\bibitem{1990255}
Chan-Gyu Lee, Yoshiaki Iijima, Tatsuhiko Hiratani, and Ken ichi Hirano.
\newblock Diffusion of chromium in \&alpha;-iron.
\newblock {\em Materials Transactions, JIM}, 31(4):255--261, 1990.

\bibitem{PhysRevB.54.11169}
G.~Kresse and J.~Furthm\"uller.
\newblock Efficient iterative schemes for ab initio total-energy calculations using a plane-wave basis set.
\newblock {\em Phys. Rev. B}, 54:11169--11186, Oct 1996.

\bibitem{PhysRevB.40.3616}
M.~Methfessel and A.~T. Paxton.
\newblock High-precision sampling for brillouin-zone integration in metals.
\newblock {\em Phys. Rev. B}, 40:3616--3621, Aug 1989.

\bibitem{lindsay2022moose}
Alexander~D. Lindsay, Derek~R. Gaston, Cody~J. Permann, Jason~M. Miller, David Andr{\v{s}}, Andrew~E. Slaughter, Fande Kong, Joshua Hansel, Robert~W. Carlsen, Casey Icenhour, Logan Harbour, Guillaume~L. Giudicelli, Roy~H. Stogner, Peter German, Jacob Badger, Sudipta Biswas, Leora Chapuis, Christopher Green, Jason Hales, Tianchen Hu, Wen Jiang, Yeon~Sang Jung, Christopher Matthews, Yinbin Miao, April Novak, John~W. Peterson, Zachary~M. Prince, Andrea Rovinelli, Sebastian Schunert, Daniel Schwen, Benjamin~W. Spencer, Swetha Veeraraghavan, Antonio Recuero, Dewen Yushu, Yaqi Wang, Andy Wilkins, and Christopher Wong.
\newblock 2.0 - {MOOSE}: Enabling massively parallel multiphysics simulation.
\newblock {\em {SoftwareX}}, 20:101202, 2022.

\bibitem{schwen2023phasefield}
D.~Schwen, L.K. Aagesen, J.W. Peterson, and M.R. Tonks.
\newblock Rapid multiphase-field model development using a modular free energy based approach with automatic differentiation in moose/marmot.
\newblock {\em Computational Materials Science}, 132:36--45, 2017.

\bibitem{KRAU202073}
T.~Krauß and S.M. Eich.
\newblock Development of a segregation model beyond mclean based on atomistic simulations.
\newblock {\em Acta Materialia}, 187:73--83, 2020.

\bibitem{WANG20211391}
Xuchao Wang, Jun Zhao, Enzhao Cui, Zhefei Sun, and Hao Yu.
\newblock Grain growth kinetics and grain refinement mechanism in {Al}2{O}3/{WC}/{TiC}/graphene ceramic composite.
\newblock {\em Journal of the European Ceramic Society}, 41(2):1391--1398, 2021.

\end{thebibliography}

\end{document}